\documentclass{article}
\usepackage{graphicx}
\usepackage{color}

\begin{document}
\title{\bf General high-order rogue waves and their dynamics in the nonlinear Schr\"odinger equation}
\author{Yasuhiro Ohta$^{1}$\footnote{Email: ohta@math.kobe-u.ac.jp}
\hspace{0.15cm}
and \hspace{0.1cm}  Jianke Yang$^{2}$\footnote{Email: jyang@math.uvm.edu} \\
{\small\it $^1$Department of Mathematics, Kobe University, Rokko,
Kobe 657-8501, Japan} \\ {\small\it $^3$Department of Mathematics
and Statistics, University of Vermont, Burlington, VT $05401$,
U.S.A}}
\date{}
\maketitle

\textbf{Abstract:} General high-order rogue waves in the nonlinear
Schr\"odinger equation are derived by the bilinear method. These
rogue waves are given in terms of determinants whose matrix elements
have simple algebraic expressions. It is shown that the general
$N$-th order rogue waves contain $N-1$ free irreducible complex
parameters. In addition, the specific rogue waves obtained by
Akhmediev et al. (Phys. Rev. E 80, 026601 (2009)) correspond to
special choices of these free parameters, and they have the highest
peak amplitudes among all rogue waves of the same order. If other
values of these free parameters are taken, however, these general
rogue waves can exhibit other solution dynamics such as arrays of
fundamental rogue waves arising at different times and spatial
positions and forming interesting patterns.

\section{Introduction}
Rogue waves, also known as freak waves, monster waves, killer waves,
extreme waves, and abnormal waves, is a hot topic in physics these
days. This name comes originally from oceanography, and it refers to
large and spontaneous ocean surface waves that occur in the sea and
are a threat even to large ships and ocean liners. Recently, an
optical analogue of rogue waves --- optical rogue waves, was
observed in optical fibres \cite{Rogue_nature1,Rogue_nature2}. These
optical rogue waves are narrow pulses which emerge from initially
weakly modulated continuous-wave signals. A growing consensus is
that both oceanic and optical rogue waves appear as a result of
modulation instability of monochromatic nonlinear waves.
Mathematically, the simplest and most universal model for the
description of modulation instability and subsequent nonlinear
evolution of quasi-monochromatic waves is the focusing nonlinear
Schr\"odinger (NLS) equation \cite{Benney,Zakharov,Hasegawa}. This
equation is integrable \cite{ZS}, thus its solutions often admit
analytical expressions. For rogue waves, the simplest (lowest-order)
analytical solution was obtained by Peregrine \cite{Rogue1}. This
solution approaches a non-zero constant background as time goes to
$\pm \infty$ but rises to a peak amplitude of three times the
background in the intermediate time. Special higher-order rogue
waves were obtained by Akhmediev, et al. using Darboux
transformation \cite{Akhmediev_PRE}. These rogue waves could reach
higher peak amplitude from a constant background. Recently, more
general higher-order (multi-Peregrine) rogue waves were obtained by
Dubard, et al. \cite{Rogue_higher_order,Rogue_higher_order2} and
Gaillard \cite{Rogue_Gaillard}. It was shown that these higher-order
waves could possess multiple intensity peaks at different points of
the space-time plane. These exact rogue-wave solutions, which sit on
non-zero constant background, are very different from the familiar
soliton and multi-soliton solutions which sit on the zero
background. These rogue waves are intimately related to homoclinic
solutions \cite{Ablowitz_homo}. Indeed, rogue waves can be obtained
from homoclinic solutions when the spatial period of homoclinic
solutions goes to infinity \cite{Rogue_homo}. These rogue waves are
also related to breather solutions which move on a non-zero constant
background with profiles changing with time
\cite{Akhmediev_breather}.

In this article, we derive general high-order rogue waves in the
nonlinear Schr\"odinger equation and explore their new solution
dynamics. Our derivation is based on the bilinear method in the
soliton theory \cite{H}. Our solution is given in terms of Gram
determinants and then further simplified so that the elements in the
determinant matrices have simple algebraic expressions. Compared to
the high-order rogue waves presented in \cite{Rogue_higher_order,
Rogue_Gaillard}, our solution appears to be more explicit and more
easily yielding specific expressions for rogue waves of any given
order. We also show that these general rogue waves of $N$-th order
contain $N-1$ free irreducible complex parameters. In addition, the
specific rogue waves obtained in \cite{Akhmediev_PRE} correspond to
special choices of these free parameters, and they have the highest
peak amplitudes among all rogue waves of the same order. If other
values of these free parameters are taken, however, these general
rogue waves can exhibit other solution dynamics such as arrays of
fundamental (Peregrine) rogue waves arising at different times and
spatial positions. Interesting patterns of these rogue-wave arrays
are also illustrated.

\section{General rogue-wave solutions}
In this article, we consider general rogue waves in the focusing NLS
equation
\begin{equation} \label{e:NLS0}
iu_t=u_{xx}+2|u|^2u.
\end{equation}
Rogue waves are nonlinear waves which approach a constant background
at large time and distances. Notice that Eq. (\ref{e:NLS0}) is
invariant under scalings $x\to \alpha x$, $t\to \alpha^2 t$, $u \to
u/\alpha$ for any real constant $\alpha$. In addition, it is
invariant under the Galilean transformation $u(x,t)\to u(x-vt, t)
\mbox{exp}(-ivx/2+iv^2t/4)$ for any real velocity $v$. Thus we only
consider rogue waves which approach unit-amplitude background at
large $x$ and $t$,
\[
u(x,t) \to e^{-2it}, \quad x, t \to \pm \infty.
\]
Then under the variable transform $u \to ue^{-2it}$, the NLS
equation (\ref{e:NLS0}) becomes
\begin{equation} \label{e:NLS}
iu_t=u_{xx}+2(|u|^2-1)u,
\end{equation}
where
\begin{equation} \label{e:bc}
u(x,t) \to 1, \quad x, t \to \pm \infty.
\end{equation}

The rogue waves are described by rational solutions in the NLS
equation. In order to present these solutions, let us introduce the
so-called elementary Schur polynomials $S_n(\mbox{\boldmath $x$})$
which are defined via the generating function,
\[
\sum_{n=0}^{\infty}S_n(\mbox{\boldmath $x$})\lambda^n
=\exp\left(\sum_{k=1}^{\infty}x_k\lambda^k\right),
\]
where $\mbox{\boldmath $x$}=(x_1,x_2,\cdots)$. For example we have
\[
S_0(\mbox{\boldmath $x$})=1, \quad S_1(\mbox{\boldmath $x$})=x_1,
\quad S_2(\mbox{\boldmath $x$})=\frac{1}{2}x_1^2+x_2, \quad
S_3(\mbox{\boldmath $x$})=\frac{1}{6}x_1^3+x_1x_2+x_3, \quad \cdots.
\]
It is known that the general Schur polynomials give the complete set
of homogeneous-weight algebraic solutions for the
Kadomtsev-Petviashvili (KP) hierarchy \cite{S,JM}.

\noindent{\bf Theorem 1.} The NLS equation (\ref{e:NLS}) under the
boundary condition (\ref{e:bc}) has nonsingular rational solutions
\begin{equation} \label{e:theorem1u}
u=\frac{\sigma_1}{\sigma_0},
\end{equation}
where
\begin{equation}   \label{e:theorem1sigma}
\sigma_n=\det_{1\le i,j\le N}\left(m_{2i-1,2j-1}^{(n)}\right),
\end{equation}
the matrix elements in $\sigma_n$ are defined by
\begin{equation}   \label{e:theorem1mij}
m_{ij}^{(n)}=\sum_{\nu=0}^{\min(i,j)}\Phi_{i\nu}^{(n)}\Psi_{j\nu}^{(n)},
\quad
\Phi_{i\nu}^{(n)}=\frac{1}{2^{\nu}}\sum_{k=0}^{i-\nu}a_k
S_{i-\nu-k}(\mbox{\boldmath $x$}^{+}(n)+\nu\mbox{\boldmath $s$}),
\quad
\Psi_{j\nu}^{(n)}=\frac{1}{2^{\nu}}\sum_{l=0}^{j-\nu}\bar a_l
S_{j-\nu-l}(\mbox{\boldmath$x$}^{-}(n)+\nu\mbox{\boldmath $s$}),
\end{equation}
$a_k$ $(k=0,1,\cdots)$ are complex constants, and $\mbox{\boldmath
$x$}^{\pm}(n)=(x^{\pm}_1(n),x^{\pm}_2,\cdots)$, $\mbox{\boldmath
$s$}=(s_1,s_2,\cdots)$ are defined by
\begin{equation} \label{e:xpm}
\begin{array}{c}
\displaystyle x^{\pm}_1(n)=x\mp 2it\pm n-\frac{1}{2}, \quad
x^{\pm}_k=\frac{x\mp 2^kit}{k!}-r_k,\quad(k\ge 2),
\\[10pt]
\displaystyle
\sum_{k=1}^{\infty}r_k\lambda^k=\ln(\cosh\frac{\lambda}{2}), \quad
\sum_{k=1}^{\infty}s_k\lambda^k=\ln(\frac{2}{\lambda}\tanh\frac{\lambda}{2}).
\end{array}
\end{equation}
The above $\sigma_n$ can also be expressed as
\begin{equation} \label{e:sigma}
\sigma_n=\sum_{\nu_1=0}^1\sum_{\nu_2=\nu_1+1}^3\sum_{\nu_3=\nu_2+1}^5\cdots
\sum_{\nu_N=\nu_{N-1}+1}^{2N-1}
\det_{1\le i,j\le N}\left(\Phi_{2i-1,\nu_j}^{(n)}\right)
\det_{1\le i,j\le N}\left(\Psi_{2i-1,\nu_j}^{(n)}\right),
\end{equation}
where we further define
\begin{equation}   \label{e:theorem1pp}
\Phi_{i\nu}^{(n)}=0,
\quad
\Psi_{i\nu}^{(n)}=0,
\quad
(i<\nu).
\end{equation}

Before deriving these rogue wave solutions in this theorem, we give
some comments. In the above definitions of $r_k$ and $s_k$, since
the generators are even functions, all odd terms are zero, i.e.,
$r_1=r_3=r_5=\cdots=0$ and $s_1=s_3=s_5=\cdots=0$. The even-term
coefficients are
\[
r_2=\frac{1}{8}, \quad r_4=-\frac{1}{192}, \quad r_6=\frac{1}{2880},
\quad \cdots, \quad s_2=-\frac{1}{12}, \quad s_4=\frac{7}{1440},
\quad s_6=-\frac{31}{90720}, \quad \cdots.
\]
In the solutions, $a_k$ are complex parameters. We will show in the
appendix that without any loss of generality, we can set
\[
a_0=1, \quad a_2=a_4=\cdots=a_{\rm even}=0.
\]
In addition, by a shift of the $x$ and $t$ axes, we can make
$a_1=0$. Thus, these solutions have $N-1$ irreducible complex
parameters, $a_3$, $a_5$, $\cdots$, $a_{2N-1}$.

\section{Derivation of general rogue-wave solutions}
In this section, we derive the general rogue-wave solutions given in
Theorem 1. This derivation utilizes the bilinear method in the
soliton theory \cite{H}. The outline of this derivation is as
follows. The NLS equation (\ref{e:NLS}) is first transformed into
the bilinear form,
\begin{equation} \label{e:blNLS}
\begin{array}{l}
(D_x^2+2)f\cdot f=2|g|^2,
\\[5pt]
(D_x^2-iD_t)g\cdot f=0,
\end{array}
\end{equation}
by the variable transformation
\begin{equation}  \label{e:ugf}
u=\frac{g}{f},
\end{equation}
where $f$ is a real variable and $g$ a complex one. Here $D$ is the
Hirota's bilinear differential operator defined by
\begin{eqnarray*}
&&P(D_x,D_y,D_t,\cdots)F(x,y,t,\cdots)\cdot G(x,y,t,\cdots)
\\
&&\quad =P(\partial_x-\partial_{x'},\partial_y-\partial_{y'},
\partial_t-\partial_{t'},\cdots)
F(x,y,t,\cdots)G(x',y',t',\cdots)|_{x'=x,y'=y,t'=t,\cdots},
\end{eqnarray*}
where $P$ is a polynomial of $D_x$, $D_y$, $D_t$, $\cdots$. Then we
consider a 2+1 dimensional generalization of the above bilinear
equation,
\begin{equation} \label{e:blDS}
\begin{array}{l}
(D_xD_y+2)f\cdot f=2gh,
\\[5pt]
(D_x^2-iD_t)g\cdot f=0,
\end{array}
\end{equation}
where $h$ is another complex variable. This is in fact the bilinear
form of the Davey-Stewartson equation, which is a 2+1 dimensional
generalization of the NLS equation. We first construct a wide class
of solutions for Eq. (\ref{e:blDS}) in the form of Gram
determinants. If the solutions $f$, $g$ and $h$ of Eq.
(\ref{e:blDS}) further satisfy the conditions,
\begin{equation} \label{e:reduction}
(\partial_x-\partial_y)f=Cf,
\end{equation}
\begin{equation} \label{e:cc}
f:{\rm real}, \quad h=\bar g,
\end{equation}
where $C$ is a constant and the overbar $\ \bar{}\ $ represents
complex conjugation, then these solutions also satisfy the bilinear
NLS equation (\ref{e:blNLS}). Among the determinant solutions for
the 2+1 dimensional system (\ref{e:blDS}), we extract algebraic
solutions satisfying the reduction condition (\ref{e:reduction}).
Then such algebraic solutions satisfy both (\ref{e:blDS}) and
(\ref{e:reduction}), i.e., they are solutions for the 1+1
dimensional system,
\begin{equation} \label{e:blpreNLS}
\begin{array}{l}
(D_x^2+2)f\cdot f=2gh,
\\[5pt]
(D_x^2-iD_t)g\cdot f=0.
\end{array}
\end{equation}
Finally we impose the real and complex conjugate condition
(\ref{e:cc}) on the algebraic solutions. Then the bilinear system
(\ref{e:blpreNLS}) reduces to the bilinear NLS equation
(\ref{e:blNLS}), hence Eq. (\ref{e:ugf}) gives the general
high-order rogue wave solutions for the NLS equation (\ref{e:NLS}).

The execution of the above derivation will involve some novel
techniques which are uncommon in the bilinear solution method
\cite{H}. It is known that the bilinear equations of the NLS
hierarchy admit homogeneous-weight polynomial solutions given by the
Schur polynomials associated with rectangular Young diagrams
\cite{IY}. However those solutions do not satisfy the complex
conjugation condition $h=\bar g$ in general, since the Schur
polynomials $g$ and $h$ in \cite{IY} have different degrees unless
the Young diagram associated with $f$ is a square. In the case of a
square-shape Young diagram for $f$, $h$ can be equal to $-\bar g$
(but not $\bar g$) and the equation becomes the defocusing NLS
equation. To construct rational solutions for the focusing NLS
equation (\ref{e:blNLS}), it is crucial to consider
weight-inhomogeneous polynomials. In order to satisfy the reduction
condition (\ref{e:reduction}) as well as the complex conjugate
condition (\ref{e:cc}), we need a specific combination of Schur
polynomials as given in Theorem 1.

Next, we follow the above outline to derive general rogue-wave
solutions to the NLS equation (\ref{e:NLS}) under the boundary
condition (\ref{e:bc}).

\subsection{Gram determinant solution for the 2+1 dimensional system}

In this subsection, we first derive the Gram determinant solution
for the 2+1 dimensional bilinear equations (\ref{e:blDS}).

\noindent{\bf Lemma 1.} Let $m_{ij}^{(n)}$, $\varphi_i^{(n)}$ and
$\psi_j^{(n)}$ be functions of $x_1$, $x_2$ and $x_{-1}$ satisfying
the following differential and difference relations,
\begin{equation} \label{e:diffrel}
\begin{array}{l}
\partial_{x_1}m_{ij}^{(n)}=\varphi_i^{(n)}\psi_j^{(n)},
\\[5pt]
\partial_{x_2}m_{ij}^{(n)}
=\varphi_i^{(n+1)}\psi_j^{(n)}+\varphi_i^{(n)}\psi_j^{(n-1)},
\\[5pt]
\partial_{x_{-1}}m_{ij}^{(n)}=-\varphi_i^{(n-1)}\psi_j^{(n+1)},
\\[5pt]
m_{ij}^{(n+1)}=m_{ij}^{(n)}+\varphi_i^{(n)}\psi_j^{(n+1)},
\\[5pt]
\partial_{x_k}\varphi_i^{(n)}=\varphi_i^{(n+k)},
\quad
\partial_{x_k}\psi_j^{(n)}=-\psi_j^{(n-k)},
\quad (k=1,2,-1).
\end{array}
\end{equation}
Then the determinant,
\begin{equation} \label{e:tauDS}
\tau_n=\det_{1\le i,j\le N}\left(m_{ij}^{(n)}\right),
\end{equation}
satisfies the bilinear equations,
\begin{equation} \label{e:bllem1}
\begin{array}{l}
(D_{x_1}D_{x_{-1}}-2)\tau_n\cdot\tau_n=-2\tau_{n+1}\tau_{n-1},
\\[5pt]
(D_{x_1}^2-D_{x_2})\tau_{n+1}\cdot\tau_n=0.
\end{array}
\end{equation}

\noindent{\bf Proof.} We have the differential formula of
determinant,
\begin{equation}  \label{e:diffdet}
\partial_x\det_{1\le i,j\le N}(a_{ij})
=\sum_{i,j=1}^N\Delta_{ij}\partial_xa_{ij},
\end{equation}
and the expansion formula of bordered determinant,
\[
\det\pmatrix{ a_{ij} &b_i \cr c_j &d}
=-\sum_{i,j}\Delta_{ij}b_ic_j+d\det(a_{ij}),
\]
where $\Delta_{ij}$ is the $(i,j)$-cofactor of the matrix
$(a_{ij})$. By using these formulae repeatedly, we can verify that
the derivatives and shifts of the $\tau$ function (\ref{e:tauDS})
are expressed by the bordered determinants as follows,
\begin{eqnarray*}
&&\partial_{x_1}\tau_n=\left|\matrix{ m_{ij}^{(n)} &\varphi_i^{(n)}
\cr -\psi_j^{(n)} &0}\right|,
\\
&&\partial_{x_1}^2\tau_n=\left|\matrix{ m_{ij}^{(n)}
&\varphi_i^{(n+1)} \cr -\psi_j^{(n)} &0}\right| +\left|\matrix{
m_{ij}^{(n)} &\varphi_i^{(n)} \cr \psi_j^{(n-1)} &0}\right|,
\\
&&\partial_{x_2}\tau_n=\left|\matrix{ m_{ij}^{(n)}
&\varphi_i^{(n+1)} \cr -\psi_j^{(n)} &0}\right| -\left|\matrix{
m_{ij}^{(n)} &\varphi_i^{(n)} \cr \psi_j^{(n-1)} &0}\right|,
\\
&&\partial_{x_{-1}}\tau_n=\left|\matrix{ m_{ij}^{(n)}
&\varphi_i^{(n-1)} \cr \psi_j^{(n+1)} &0}\right|,
\\
&&(\partial_{x_1}\partial_{x_{-1}}-1)\tau_n=\left|\matrix{
m_{ij}^{(n)} &\varphi_i^{(n-1)} &\varphi_i^{(n)} \cr \psi_j^{(n+1)}
&0 &-1 \cr -\psi_j^{(n)} &-1 &0}\right|,
\\
&&\tau_{n+1}=\left|\matrix{ m_{ij}^{(n)} &\varphi_i^{(n)} \cr
-\psi_j^{(n+1)} &1}\right|,
\\
&&\tau_{n-1}=\left|\matrix{ m_{ij}^{(n)} &\varphi_i^{(n-1)} \cr
\psi_j^{(n)} &1}\right|,
\\
&&\partial_{x_1}\tau_{n+1}=\left|\matrix{ m_{ij}^{(n)}
&\varphi_i^{(n+1)} \cr -\psi_j^{(n+1)} &0}\right|,
\\
&&\partial_{x_1}^2\tau_{n+1}=\left|\matrix{ m_{ij}^{(n)}
&\varphi_i^{(n+2)} \cr -\psi_j^{(n+1)} &0}\right| +\left|\matrix{
m_{ij}^{(n)} &\varphi_i^{(n)} &\varphi_i^{(n+1)} \cr -\psi_j^{(n)}
&0 &0 \cr -\psi_j^{(n+1)} &1 &0}\right|,
\\
&&\partial_{x_2}\tau_{n+1}=\left|\matrix{ m_{ij}^{(n)}
&\varphi_i^{(n+2)} \cr -\psi_j^{(n+1)} &0}\right| -\left|\matrix{
m_{ij}^{(n)} &\varphi_i^{(n)} &\varphi_i^{(n+1)} \cr -\psi_j^{(n)}
&0 &0 \cr -\psi_j^{(n+1)} &1 &0}\right|.
\end{eqnarray*}
{}From the Jacobi formula of determinants,
\[
\left|\matrix{ a_{ij} &b_i &c_i \cr d_j &e &f \cr g_j &h &k}\right|
\times\left|\matrix{ a_{ij}}\right| =\left|\matrix{ a_{ij} &c_i \cr
g_j &k}\right| \times\left|\matrix{ a_{ij} &b_i \cr d_j &e}\right|
-\left|\matrix{ a_{ij} &b_i \cr g_j &h}\right| \times\left|\matrix{
a_{ij} &c_i \cr d_j &f}\right|,
\]
we immediately obtain the identities,
\begin{eqnarray*}
&&(\partial_{x_1}\partial_{x_{-1}}-1)\tau_n\times\tau_n
=\partial_{x_1}\tau_n\times\partial_{x_{-1}}\tau_n
-(-\tau_{n-1})(-\tau_{n+1}),
\\
&&\frac{1}{2}(\partial_{x_1}^2-\partial_{x_2})\tau_{n+1}\times\tau_n
=\partial_{x_1}\tau_{n+1}\times\partial_{x_1}\tau_n
-\tau_{n+1}\frac{1}{2}(\partial_{x_1}^2+\partial_{x_2})\tau_n,
\end{eqnarray*}
which are the bilinear equations (\ref{e:bllem1}). This completes
the proof. \hfill\fbox{}

Since the matrix element $m_{ij}^{(n)}$ is written as
\[
m_{ij}^{(n)}=\int^{x_1}\varphi_i^{(n)}\psi_j^{(n)}dx_1,
\]
the determinant (\ref{e:tauDS}) is often called the Gram determinant
solution. Let us define
\[
f=\tau_0, \quad g=\tau_1, \quad h=\tau_{-1},
\]
then these are the Gram determinant solution for the 2+1 dimensional
system,
\begin{eqnarray*}
&&(D_{x_1}D_{x_{-1}}-2)f\cdot f=-2gh,
\\
&&(D_{x_1}^2-D_{x_2})g\cdot f=0,
\end{eqnarray*}
which is nothing but the bilinear equations (\ref{e:blDS}) by
writing $x_1=x$, $x_2=-it$ and $x_{-1}=-y$.

\subsection{Algebraic solutions for the 1+1 dimensional system}

Next we derive algebraic solutions satisfying both the bilinear
equations (\ref{e:blDS}) and the reduction condition
(\ref{e:reduction}), hence satisfying the 1+1 dimensional system
(\ref{e:blpreNLS}). These solutions are obtained by choosing the
matrix elements appropriately in the Gram determinant solution in
Lemma 1.

\noindent{\bf Lemma 2.} We define matrix elements $m_{ij}^{(n)}$ by
\begin{equation} \label{e:mij}
m_{ij}^{(n)} =\left.A_iB_jm^{(n)}\right|_{p=1,q=1},
\end{equation}
\begin{equation} \label{e:mn}
m^{(n)}=\frac{1}{p+q}(-\frac{p}{q})^ne^{\xi+\eta}, \quad
\xi=px_1+p^2x_2, \quad \eta=qx_1-q^2x_2,
\end{equation}
where $A_i$ and $B_j$ are differential operators with respect to $p$
and $q$ respectively, defined as
\begin{eqnarray*}
&&A_0=a_0,
\\
&&A_1=a_0p\partial_p+a_1,
\\
&&A_2=\frac{a_0}{2}(p\partial_p)^2+a_1p\partial_p+a_2,
\\
&&\quad\vdots
\\
&&A_i=\sum_{k=0}^i\frac{a_k}{(i-k)!}(p\partial_p)^{i-k},
\end{eqnarray*}
and
\begin{eqnarray*}
&&B_0=b_0,
\\
&&B_1=b_0q\partial_q+b_1,
\\
&&B_2=\frac{b_0}{2}(q\partial_q)^2+b_1q\partial_q+b_2,
\\
&&\quad\vdots
\\
&&B_j=\sum_{l=0}^j\frac{b_l}{(j-l)!}(q\partial_q)^{j-l},
\end{eqnarray*}
and $a_k$ and $b_l$ are constants. Then the determinant
\begin{equation} \label{e:tau}
\tau_n=\det_{1\le i,j\le
N}\left(m_{2i-1,2j-1}^{(n)}\right)=\left|\matrix{ m_{11}^{(n)}
&m_{13}^{(n)} &\cdots &m_{1,2N-1}^{(n)} \cr m_{31}^{(n)}
&m_{33}^{(n)} &\cdots &m_{3,2N-1}^{(n)} \cr \vdots &\vdots &&\vdots
\cr m_{2N-1,1}^{(n)} &m_{2N-1,3}^{(n)} &\cdots
&m_{2N-1,2N-1}^{(n)}}\right|
\end{equation}
satisfies the bilinear equations
\begin{equation}  \label{e:bilinearlemma2}
\begin{array}{l}
(D_{x_1}^2+2)\tau_n\cdot\tau_n=2\tau_{n+1}\tau_{n-1},
\\[5pt]
(D_{x_1}^2-D_{x_2})\tau_{n+1}\cdot\tau_n=0.
\end{array}
\end{equation}

\noindent{\bf Proof.} First let us introduce $\tilde m^{(n)}$,
$\tilde\varphi^{(n)}$ and $\tilde\psi^{(n)}$ by
\[
\tilde
m^{(n)}=\frac{1}{p+q}(-\frac{p}{q})^ne^{\tilde\xi+\tilde\eta}, \quad
\tilde\varphi^{(n)}=p^ne^{\tilde\xi}, \quad
\tilde\psi^{(n)}=(-q)^{-n}e^{\tilde\eta},
\]
where
\[
\tilde\xi=\frac{1}{p}x_{-1}+px_1+p^2x_2, \quad
\tilde\eta=\frac{1}{q}x_{-1}+qx_1-q^2x_2.
\]
Obviously these functions satisfy the differential and difference
relations
\begin{eqnarray*}
&&\partial_{x_1}\tilde m^{(n)}=\tilde\varphi^{(n)}\tilde\psi^{(n)},
\\
&&\partial_{x_2}\tilde m^{(n)}
=\tilde\varphi^{(n+1)}\tilde\psi^{(n)}+\tilde\varphi^{(n)}\tilde\psi^{(n-1)},
\\
&&\partial_{x_{-1}}\tilde
m^{(n)}=-\tilde\varphi^{(n-1)}\tilde\psi^{(n+1)},
\\
&&\tilde m^{(n+1)}=\tilde
m^{(n)}+\tilde\varphi^{(n)}\tilde\psi^{(n+1)},
\\
&&\partial_{x_k}\tilde\varphi^{(n)}=\tilde\varphi^{(n+k)}, \quad
\partial_{x_k}\tilde\psi^{(n)}=-\tilde\psi^{(n-k)},
\quad (k=1,2,-1).
\end{eqnarray*}
Therefore, by defining
\[
\tilde m_{ij}^{(n)}=A_iB_j\tilde m^{(n)}, \quad
\tilde\varphi_i^{(n)}=A_i\tilde\varphi^{(n)}, \quad
\tilde\psi_j^{(n)}=B_j\tilde\psi^{(n)},
\]
we see that these $\tilde m_{ij}^{(n)}$, $\tilde\varphi_i^{(n)}$ and
$\tilde\psi_j^{(n)}$ obey the differential and difference relations
(\ref{e:diffrel}) since the operators $A_i$ and $B_j$ commute with
differentials $\partial_{x_k}$. Lemma 1 then tells us that for an
arbitrary sequence of indices
$(i_1,i_2,\cdots,i_N;j_1,j_2,\cdots,j_N)$, the determinant
\[
\tilde\tau_n=\det_{1\le\nu,\mu\le N}\left(\tilde
m_{i_\nu,j_\mu}^{(n)}\right)
\]
satisfies the bilinear equations (\ref{e:bllem1}). For example,
\[
\tilde\tau_n=\det_{1\le i,j\le N}\left(\tilde
m_{2i-1,2j-1}^{(n)}\right),
\]
is a solution to Eq. (\ref{e:bllem1}), where $p$ and $q$ are
arbitrary parameters.

Next we consider the reduction condition.
{}From the Leibniz rule, we have the operator relation,
\[
(p\partial_p)^m(p+\frac{1}{p})=\sum_{l=0}^m{m \choose l}
(p+(-1)^l\frac{1}{p})(p\partial_p)^{m-l},
\]
thus we get
\begin{eqnarray*}
&&A_i(p+\frac{1}{p})=\sum_{k=0}^i\frac{a_k}{(i-k)!}
\sum_{l=0}^{i-k}{i-k \choose
l}(p+(-1)^l\frac{1}{p})(p\partial_p)^{i-k-l}
\\
&&\quad =\sum_{l=0}^{i}\sum_{k=0}^{i-l}
\frac{a_k}{l!\,(i-k-l)!}(p+(-1)^l\frac{1}{p})(p\partial_p)^{i-k-l}
=\sum_{l=0}^{i}\frac{1}{l!}(p+(-1)^l\frac{1}{p})A_{i-l},
\end{eqnarray*}
and similarly
\[
B_j(q+\frac{1}{q})=\sum_{l=0}^{j}\frac{1}{l!}(q+(-1)^l\frac{1}{q})B_{j-l}.
\]
By using these relations, we find that $\tilde m_{ij}^{(n)}$
satisfies
\begin{eqnarray*}
&&(\partial_{x_1}+\partial_{x_{-1}})\tilde m_{ij}^{(n)}
=A_iB_j(\partial_{x_1}+\partial_{x_{-1}})\tilde m^{(n)}
=A_iB_j(p+q+\frac{1}{p}+\frac{1}{q})\tilde m^{(n)}
\\
&&\quad
=\sum_{k=0}^{i}\frac{1}{k!}(p+(-1)^k\frac{1}{p})A_{i-k}B_j\tilde
m^{(n)}
+\sum_{l=0}^{j}\frac{1}{l!}(q+(-1)^l\frac{1}{q})A_iB_{j-l}\tilde
m^{(n)}
\\
&&\quad =\sum_{k=0}^{i}\frac{1}{k!}(p+(-1)^k\frac{1}{p})\tilde
m_{i-k,j}^{(n)}
+\sum_{l=0}^{j}\frac{1}{l!}(q+(-1)^l\frac{1}{q})\tilde
m_{i,j-l}^{(n)}.
\end{eqnarray*}
Now let us take $p=1$ and $q=1$. Then $\left.\tilde
m_{ij}^{(n)}\right|_{p=1,q=1}$ satisfies the contiguity relation,
\begin{eqnarray} \label{e:contig}
&&(\partial_{x_1}+\partial_{x_{-1}}) \left(\left.\tilde
m_{ij}^{(n)}\right|_{p=1,q=1}\right) =2\sum_{\scriptstyle k=0 \atop
\scriptstyle k:{\rm even}}^{i} \frac{1}{k!}\left.\tilde
m_{i-k,j}^{(n)}\right|_{p=1,q=1} +2\sum_{\scriptstyle l=0 \atop
\scriptstyle l:{\rm even}}^{j} \frac{1}{l!}\left.\tilde
m_{i,j-l}^{(n)}\right|_{p=1,q=1}.
\end{eqnarray}
By using the formula (\ref{e:diffdet}) and the above relation, the
differential of the determinant,
\[
\tilde{\tilde\tau}_n=\det_{1\le i,j\le N} \left(\left.\tilde
m_{2i-1,2j-1}^{(n)}\right|_{p=1,q=1}\right)
\]
is calculated as
\begin{eqnarray*}
&&(\partial_{x_1}+\partial_{x_{-1}})\tilde{\tilde\tau}_n
=\sum_{i=1}^N\sum_{j=1}^N\Delta_{ij}(\partial_{x_1}+\partial_{x_{-1}})
\left(\left.\tilde m_{2i-1,2j-1}^{(n)}\right|_{p=1,q=1}\right)
\\
&&\quad =\sum_{i=1}^N\sum_{j=1}^N\Delta_{ij}
\left(2\sum_{\scriptstyle k=0 \atop \scriptstyle k:{\rm
even}}^{2i-1} \frac{1}{k!}\left.\tilde
m_{2i-1-k,2j-1}^{(n)}\right|_{p=1,q=1} +2\sum_{\scriptstyle l=0
\atop \scriptstyle l:{\rm even}}^{2j-1} \frac{1}{l!}\left.\tilde
m_{2i-1,2j-1-l}^{(n)}\right|_{p=1,q=1}\right)
\\
&&\quad =2\sum_{i=1}^N \sum_{\scriptstyle k=0 \atop \scriptstyle
k:{\rm even}}^{2i-1} \frac{1}{k!}\sum_{j=1}^N\Delta_{ij}
\left.\tilde m_{2i-1-k,2j-1}^{(n)}\right|_{p=1,q=1} +2\sum_{j=1}^N
\sum_{\scriptstyle l=0 \atop \scriptstyle l:{\rm even}}^{2j-1}
\frac{1}{l!}\sum_{i=1}^N\Delta_{ij} \left.\tilde
m_{2i-1,2j-1-l}^{(n)}\right|_{p=1,q=1},
\end{eqnarray*}
where $\Delta_{ij}$ is the $(i,j)$-cofactor of
$\displaystyle\mathop{\rm mat}_{1\le i,j\le N} \left(\left.\tilde
m_{2i-1,2j-1}^{(n)}\right|_{p=1,q=1}\right)$. In the first term of
the right-hand side, only the term with $k=0$ survives and the other
terms vanish, since for $k=2,4,\cdots$, the summation with respect
to $j$ is a determinant with two identical rows. Similarly in the
second term, only the term with $l=0$ remains. Thus the right side
of the above equation becomes
\[
2\sum_{i=1}^N \sum_{j=1}^N\Delta_{ij}\left.\tilde
m_{2i-1,2j-1}^{(n)}\right|_{p=1,q=1} +2\sum_{j=1}^N
\sum_{i=1}^N\Delta_{ij}\left.\tilde
m_{2i-1,2j-1}^{(n)}\right|_{p=1,q=1} =4N\tilde{\tilde\tau}_n.
\]
Therefore $\tilde{\tilde\tau}_n$ satisfies the reduction condition
\begin{equation} \label{e:redcond}
(\partial_{x_1}+\partial_{x_{-1}})\tilde{\tilde\tau}_n=4N\tilde{\tilde\tau}_n.
\end{equation}
Since $\tilde{\tilde\tau}_n$ is a special case of $\tilde\tau_n$, it
also satisfies the bilinear equations (\ref{e:bllem1}) with $\tau_n$
replaced by $\tilde{\tilde\tau}_n$. From (\ref{e:bllem1}) and
(\ref{e:redcond}), we see that $\tilde{\tilde\tau}_n$ satisfies the
1+1 dimensional bilinear equations
\begin{eqnarray*}
&&(D_{x_1}^2+2)\tilde{\tilde\tau}_n\cdot\tilde{\tilde\tau}_n
=2\tilde{\tilde\tau}_{n+1}\tilde{\tilde\tau}_{n-1},
\\
&&(D_{x_1}^2-D_{x_2})\tilde{\tilde\tau}_{n+1}\cdot\tilde{\tilde\tau}_n=0,
\end{eqnarray*}
which are the same as Eq. (\ref{e:bilinearlemma2}). Now we can take
$x_{-1}=0$, then $\left.\tilde m_{ij}^{(n)}\right|_{p=1,q=1}$ and
$\tilde{\tilde\tau}_n$ reduce to $m_{ij}^{(n)}$ and $\tau_n$ in
Lemma 2, and this $\tau_n$ satisfies the bilinear equations
(\ref{e:bilinearlemma2}). This completes the proof. \hfill\fbox{}

The above proof uses the technique of reduction. The reduction is a
procedure to derive solutions of a lower dimensional system from
those of a higher dimensional one. By using the reduction condition
(\ref{e:redcond}), the derivative with respect to a variable
$x_{-1}$ is replaced by the derivative with respect to another
variable $x_1$. Then in the solution, $x_{-1}$ is just a parameter
to which we can substitute any value (such as zero as we did above).

It is remarkable that the determinant expression of the solution
(\ref{e:tau}) has a quite unique structure: the indices of matrix
elements, which label the degree of polynomial, have the step of 2.
This comes from the requirement of the reduction condition, i.e.,
since the contiguity relation (\ref{e:contig}) relates matrix
elements with indices shifted by even numbers, we want such a
determinant to satisfy the reduction condition. This type of Gram
determinant solutions has not been reported in the literature to the
best of the authors' knowledge.

{}From Lemma 2, by writing $x_1=x$ and $x_2=-it$, we find that
$f=\tau_0$, $g=\tau_1$ and $h=\tau_{-1}$ satisfy the 1+1 dimensional
system (\ref{e:blpreNLS}).

\subsection{Complex conjugacy and regularity}

Now we consider the complex conjugate condition (\ref{e:cc}) and the
regularity (nonsingularity) of solutions. This complex conjugate
condition now is
\[
\tau_0:{\rm real}, \quad \tau_{-1}=\bar\tau_1.
\]
Since $x_1=x$ is real and $x_2=-it$ is pure imaginary in Lemma 2,
the above condition is easily satisfied by taking the parameters
$a_k$ and $b_k$ to be complex conjugate to each other,
\begin{equation} \label{e:ccab}
b_k=\bar a_k.
\end{equation}
In fact, under the condition (\ref{e:ccab}) we have
\[
\overline{m_{ij}^{(n)}}
=\left.m_{ij}^{(n)}\right|_{a_k\leftrightarrow
b_k,x_2\leftrightarrow -x_2} =m_{ji}^{(-n)},
\]
and therefore
\[
\bar\tau_n=\tau_{-n}.
\]

Under condition (\ref{e:ccab}), we can further show that the
rational solution $u=g/f=\tau_1/\tau_0$ is nonsingular, i.e.,
$\tau_0$ is nonzero for all $(x,t)$. To prove it, we notice that
$f=\tau_0$ is the determinant of a Hermitian matrix $\displaystyle
M=\mathop{\rm mat}_{1\le i,j\le N}\left(m_{2i-1,2j-1}^{(0)}\right)$.
For any non-zero column vector $\mbox{\boldmath
$v$}=(v_1,v_2,\cdots,v_N)^T$ and $\bar{\mbox{\boldmath $v$}}$ being
its complex transpose, we have
\begin{eqnarray*}
&&\bar{\mbox{\boldmath $v$}}M\mbox{\boldmath $v$}
=\sum_{i,j=1}^N\bar v_im_{2i-1,2j-1}^{(0)}v_j =\sum_{i,j=1}^N\bar
v_iv_j
\left.A_{2i-1}B_{2j-1}\frac{1}{p+q}e^{\xi+\eta}\right|_{p=1,q=1}
\\
&&\quad =\sum_{i,j=1}^N\bar v_iv_j
\left.A_{2i-1}B_{2j-1}\int_{-\infty}^xe^{\xi+\eta}dx\right|_{p=1,q=1}
=\int_{-\infty}^x\sum_{i,j=1}^N\left.\bar v_iv_j
A_{2i-1}B_{2j-1}e^{\xi+\eta}\right|_{p=1,q=1}dx
\\
&&\quad =\int_{-\infty}^x\left|\sum_{i=1}^N\left.\bar v_i
A_{2i-1}e^{\xi}\right|_{p=1}\right|^2dx>0,
\end{eqnarray*}
which proves that the Hermitian matrix $M$ is positive definite.
Therefore the denominator $f=\det M>0$, so the solution $u$ is
nonsingular.

It is noted that the above proofs of complex conjugate condition and
regularity condition are quite easy. This is an advantage of the
Gram determinant expression of solutions (as compared to the
Wronskian expression).

Summarizing the above results, we obtain the following intermediate
theorem on rogue-wave solutions in the NLS equation.

\noindent{\bf Theorem 2.} The NLS equation (\ref{e:NLS}) has the
nonsingular rational solutions,
\begin{equation}  \label{e:uTheorem2}
u=\frac{\tau_1}{\tau_0},
\end{equation}
where
\begin{equation}
\tau_n=\det_{1\le i,j\le N}\left(m_{2i-1,2j-1}^{(n)}\right),
\end{equation}
where the matrix elements are defined by
\begin{equation} \label{e:q4}
m_{ij}^{(n)}=\sum_{k=0}^i\sum_{l=0}^j \frac{a_k}{(i-k)!}\frac{\bar
a_l}{(j-l)!} \left.(p\partial_p)^{i-k}(q\partial_q)^{j-l}
\frac{1}{p+q}(-\frac{p}{q})^ne^{(p+q)x-(p^2-q^2)\sqrt{-1}\hspace{0.06cm}
t}\right|_{p=1,q=1},
\end{equation}
and $a_k$ are complex constants.

\subsection{Simplification of rogue-wave solutions}

Finally we simplify the rogue-wave solutions in Theorem 2 and derive
the solution formulae given in Theorem 1. The generator ${\cal G}$
of the differential operators $(p\partial_p)^k(q\partial_q)^l$ is
given as
\[
{\cal G}=\sum_{k=0}^{\infty}\sum_{l=0}^{\infty}
\frac{\kappa^k}{k!}\frac{\lambda^l}{l!}(p\partial_p)^k(q\partial_q)^l
=\exp(\kappa p\partial_p+\lambda q\partial_q)
=\exp(\kappa\partial_{\ln p}+\lambda\partial_{\ln q}),
\]
thus for any function $F(p,q)$, we have
\begin{equation} \label{r:GFpq}
{\cal G}F(p,q)=F(e^{\kappa}p,e^{\lambda}q).
\end{equation}
This relation can also be seen by expanding its right hand side into
Taylor series of $(\kappa, \lambda)$ around the point $(0,0)$. By
applying this relation to
\[
m^{(n)}=\frac{1}{p+q}(-\frac{p}{q})^n\exp\left((p+q)x-(p^2-q^2)it\right),
\]
we get
\[
{\cal G}m^{(n)}
=\frac{1}{e^{\kappa}p+e^{\lambda}q}(-\frac{e^{\kappa}p}{e^{\lambda}q})^n
\exp\left((e^{\kappa}p+e^{\lambda}q)x
-(e^{2\kappa}p^2-e^{2\lambda}q^2)it\right),
\]
thus
\begin{eqnarray*}
&&\left.\frac{1}{m^{(n)}}{\cal G}m^{(n)}\right|_{p=1,q=1}
=\frac{2}{e^{\kappa}+e^{\lambda}}e^{n(\kappa-\lambda)}
\exp\left((e^{\kappa}+e^{\lambda}-2)x-(e^{2\kappa}-e^{2\lambda})it\right)
\\
&&\quad =\frac{1}
{1-\frac{(e^{\kappa}-1)(e^{\lambda}-1)}{(e^{\kappa}+1)(e^{\lambda}+1)}}
\exp\left(n(\kappa-\lambda)
+(e^{\kappa}+e^{\lambda}-2)x-(e^{2\kappa}-e^{2\lambda})it
-\ln\frac{(e^{\kappa}+1)(e^{\lambda}+1)}{4}\right).
\end{eqnarray*}
In the most right-hand side, the exponent is rewritten as
\[
n(\kappa-\lambda) +\sum_{k=1}^{\infty}\frac{\kappa^k}{k!}(x-2^kit)
+\sum_{l=1}^{\infty}\frac{\lambda^l}{l!}(x+2^lit)
-\frac{\kappa}{2}-\frac{\lambda}{2}
-\ln\left(\cosh\frac{\kappa}{2}\cosh\frac{\lambda}{2}\right)
=\sum_{k=1}^{\infty}x^{+}_k\kappa^k+\sum_{l=1}^{\infty}x^{-}_l\lambda^l,
\]
where $x^{+}_k$ and $x^{-}_l$ are defined in (\ref{e:xpm}), and the
prefactor is rewritten as
\[
\sum_{\nu=0}^{\infty}\left(\frac{(e^{\kappa}-1)(e^{\lambda}-1)}
{(e^{\kappa}+1)(e^{\lambda}+1)}\right)^{\nu}
=\sum_{\nu=0}^{\infty}\left(\frac{\kappa\lambda}{4}\right)^{\nu}
\exp\left(\nu\ln\left(\frac{4}{\kappa\lambda}
\tanh\frac{\kappa}{2}\tanh\frac{\lambda}{2}\right)\right)
=\sum_{\nu=0}^{\infty}\left(\frac{\kappa\lambda}{4}\right)^{\nu}
\exp\left(\nu\sum_{k=1}^{\infty}s_k(\kappa^k+\lambda^k)\right),
\]
where $s_k$ is defined in (\ref{e:xpm}). Therefore we obtain
\[
\left.\frac{1}{m^{(n)}}{\cal G}m^{(n)}\right|_{p=1,q=1}
=\sum_{\nu=0}^{\infty}\left(\frac{\kappa\lambda}{4}\right)^{\nu}
\exp\left(\sum_{k=1}^{\infty}(x^{+}_k+\nu s_k)\kappa^k
+\sum_{l=1}^{\infty}(x^{-}_l+\nu s_l)\lambda^l\right),
\]
and taking the coefficient of $\kappa^k\lambda^l$ of both sides, we
find
\[
\left.\frac{1}{m^{(n)}}\frac{1}{k!l!}(p\partial_p)^k(q\partial_q)^l
m^{(n)}\right|_{p=1,q=1} =\sum_{\nu=0}^{\min(k,l)}\frac{1}{4^{\nu}}
S_{k-\nu}(\mbox{\boldmath $x$}^{+}+\nu\mbox{\boldmath $s$})
S_{l-\nu}(\mbox{\boldmath $x$}^{-}+\nu\mbox{\boldmath $s$}).
\]
Using the above results, the matrix element of the Gram determinant
is then calculated as
\begin{eqnarray*}
&& \left.\frac{1}{m^{(n)}}A_iB_jm^{(n)}\right|_{p=1,q=1}
=\sum_{k=0}^i\sum_{l=0}^ja_k\bar a_l
\sum_{\nu=0}^{\min(i-k,j-l)}\frac{1}{4^{\nu}}
S_{i-k-\nu}(\mbox{\boldmath $x$}^{+}+\nu\mbox{\boldmath $s$})
S_{j-l-\nu}(\mbox{\boldmath $x$}^{-}+\nu\mbox{\boldmath $s$})
\\
&&\quad =\sum_{\nu=0}^{\min(i,j)}\frac{1}{4^{\nu}}
\sum_{k=0}^{i-\nu}\sum_{l=0}^{j-\nu}a_k\bar a_l
S_{i-k-\nu}(\mbox{\boldmath $x$}^{+}+\nu\mbox{\boldmath $s$})
S_{j-l-\nu}(\mbox{\boldmath $x$}^{-}+\nu\mbox{\boldmath $s$}).
\end{eqnarray*}
Putting $\sigma_n=\tau_n/(\left.m^{(n)}\right|_{p=1,q=1})^N$,
we obtain the determinant expression in (\ref{e:theorem1sigma})
and (\ref{e:theorem1mij}).
Finally by using (\ref{e:theorem1pp}) and the formula,
\[
\det(a_{ij}+b_ic_j)=\det\pmatrix{ a_{ij} &b_i \cr -c_j &1},
\]
repeatedly, the determinant $\sigma_n$ can be rewritten into the following
$3N\times 3N$ determinant form,
\begin{eqnarray*}
&&\sigma_n=\det_{1\le i,j\le N}\left(\sum_{\nu=0}^{\min(2i-1,2j-1)}
\Phi_{2i-1,\nu}^{(n)}\Psi_{2j-1,\nu}^{(n)}\right)
=\det_{1\le i,j\le N}\left(\sum_{\nu=0}^{2N-1}
\Phi_{2i-1,\nu}^{(n)}\Psi_{2j-1,\nu}^{(n)}\right)
\\
&&\quad
=\left|\matrix{O &\matrix{
\Phi_{10}^{(n)} &\Phi_{11}^{(n)} &\cdots &\Phi_{1,2N-1}^{(n)} \cr
\Phi_{30}^{(n)} &\Phi_{31}^{(n)} &\cdots &\Phi_{3,2N-1}^{(n)} \cr
\vdots &\vdots &&\vdots \cr
\Phi_{2N-1,0}^{(n)} &\Phi_{2N-1,1}^{(n)} &\cdots &\Phi_{2N-1,2N-1}^{(n)}} \cr
\matrix{
-\Psi_{10}^{(n)} &-\Psi_{30}^{(n)} &\cdots &-\Psi_{2N-1,0}^{(n)} \cr
-\Psi_{11}^{(n)} &-\Psi_{31}^{(n)} &\cdots &-\Psi_{2N-1,1}^{(n)} \cr
\vdots &\vdots &&\vdots \cr
-\Psi_{1,2N-1}^{(n)} &-\Psi_{3,2N-1}^{(n)} &\cdots &-\Psi_{2N-1,2N-1}^{(n)}}
&I}\right|,
\end{eqnarray*}
where $O$ and $I$ are the $N\times N$ zero matrix and $2N\times 2N$
unit matrix, respectively.
Applying the Laplace expansion to the above determinant, we get
\[
\sigma_n=\sum_{0\le \nu_1<\nu_2<\cdots<\nu_N\le 2N-1}
\left|\matrix{
\Phi_{1\nu_1}^{(n)} &\Phi_{1\nu_2}^{(n)} &\cdots &\Phi_{1\nu_N}^{(n)} \cr
\Phi_{3\nu_1}^{(n)} &\Phi_{3\nu_2}^{(n)} &\cdots &\Phi_{3\nu_N}^{(n)} \cr
\vdots &\vdots &&\vdots \cr
\Phi_{2N-1,\nu_1}^{(n)} &\Phi_{2N-1,\nu_2}^{(n)} &\cdots
 &\Phi_{2N-1,\nu_N}^{(n)}}\right|
\times\left|\matrix{
\Psi_{1\nu_1}^{(n)} &\Psi_{3\nu_1}^{(n)} &\cdots &\Psi_{2N-1,\nu_1}^{(n)} \cr
\Psi_{1\nu_2}^{(n)} &\Psi_{3\nu_2}^{(n)} &\cdots &\Psi_{2N-1,\nu_2}^{(n)} \cr
\vdots &\vdots &&\vdots \cr
\Psi_{1\nu_N}^{(n)} &\Psi_{3\nu_N}^{(n)} &\cdots &\Psi_{2N-1,\nu_N}^{(n)}}
\right|,
\]
and noticing (\ref{e:theorem1pp}), the expanded expression (\ref{e:sigma})
is obtained. Theorem 1 is then proved.

\subsection{Boundary conditions}

In order to show the boundary asymptotics (\ref{e:bc}), let us
estimate the degree of polynomials of the denominator and numerator
in (\ref{e:theorem1u}). The elementary Schur polynomial
$S_k(\mbox{\boldmath $x$})$ has the form $S_k(\mbox{\boldmath
$x$})=(x_1)^k/k!+\mbox{(lower degree terms)}$, where
$\mbox{\boldmath $x$}=(x_1,x_2,\cdots)$. Thus the degree of the
polynomial $S_k(\mbox{\boldmath $x$}^{\pm}+\nu\mbox{\boldmath $s$})$
in $(x,t)$ is $k$ and its leading term appears in the monomial
$(x^{\pm}_1)^k/k!$, i.e., the leading term is given by $(x\mp
2it)^k/k!$. Therefore the degrees of $\Phi_{j\nu}^{(n)}$ and
$\Psi_{j\nu}^{(n)}$ are both $j-\nu$, and their leading terms are
$a_0(x-2it)^{j-\nu}/(j-\nu)!2^\nu$ and $\bar
a_0(x+2it)^{j-\nu}/(j-\nu)!2^\nu$, respectively. Therefore both of
the degrees of determinants $\displaystyle\det_{1\le i,j\le
N}\left(\Phi_{2i-1,\nu_j}^{(n)}\right)$ and $\displaystyle\det_{1\le
i,j\le N}\left(\Psi_{2i-1,\nu_j}^{(n)}\right)$ are given by
$1+3+\cdots+(2N-1)-\nu_1-\nu_2-\cdots-\nu_N$, and in the expression
(\ref{e:sigma}), the highest degree term comes from the term of
$\nu_1=0$, $\nu_2=1$, $\cdots$, $\nu_N=N-1$ in the summation. For
$\nu_j=j-1$, we have
\begin{eqnarray*}
&&\det_{1\le i,j\le N}\left(\Phi_{2i-1,j-1}^{(n)}\right)
=\left|\matrix{
a_0x^{+}_1 &\frac{a_0}{2} &0 &0 &0 &\cdots \cr
\frac{a_0(x^{+}_1)^3}{3!} &\frac{a_0(x^{+}_1)^2}{2!2} &\frac{a_0x^{+}_1}{2^2}
&\frac{a_0}{2^3} &0 &\cdots \cr
\vdots &\vdots &\vdots &\vdots \cr
\frac{a_0(x^{+}_1)^{2N-1}}{(2N-1)!} &\frac{a_0(x^{+}_1)^{2N-2}}{(2N-2)!2}
&\frac{a_0(x^{+}_1)^{2N-3}}{(2N-3)!2^2} &\frac{a_0(x^{+}_1)^{2N-4}}{(2N-4)!2^3}
&\cdots &\frac{a_0(x^{+}_1)^N}{N!2^{N-1}}}\right|+\mbox{(lower degree terms)}
\\
&&\quad
=\frac{a_0^N(x^{+}_1)^{N(N+1)/2}}{1!3!\cdots(2N-1)!2^{N(N-1)/2}}
\\
&&\quad
\times\left|\matrix{
1 &1 &0 &0 &0 &\cdots \cr
1 &3 &3\cdot 2 &3\cdot 2\cdot 1 &0 &\cdots \cr
\vdots &\vdots &\vdots &\vdots \cr
1 &2N-1 &(2N-1)(2N-2) &(2N-1)(2N-2)(2N-3)
&\cdots &(2N-1)(2N-2)\cdots(N+1)}\right|
\\
&&\quad
+\mbox{(lower degree terms)}.
\end{eqnarray*}
The above determinant is equal to
\[
\displaystyle\det_{1\le i,j\le
N}\left(\prod_{\nu=1}^{j-1}(2i-\nu)\right) =\det_{1\le i,j\le
N}\left((2i-1)^{j-1}\right),
\]
which is the Vandermonde
determinant. Thus we obtain
\[
\det_{1\le i,j\le N}\left(\Phi_{2i-1,j-1}^{(n)}\right)
=\frac{0!1!\cdots(N-1)!}{1!3!\cdots(2N-1)!}a_0^N(x^{+}_1)^{N(N+1)/2}
+\mbox{(lower degree terms)},
\]
and similarly
\[
\det_{1\le i,j\le N}\left(\Psi_{2i-1,j-1}^{(n)}\right)
=\frac{0!1!\cdots(N-1)!}{1!3!\cdots(2N-1)!}\bar a_0^N(x^{-}_1)^{N(N+1)/2}
+\mbox{(lower degree terms)}.
\]
Consequently, the leading term of $\sigma_n$ is given by
\[
\left(\frac{0!1!\cdots(N-1)!}{1!3!\cdots(2N-1)!}\right)^2
|a_0|^{2N}(x^2+4t^2)^{N(N+1)/2},
\]
which is independent of $n$. Hence $u=\sigma_1/\sigma_0$ satisfies
the boundary condition (\ref{e:bc}).

\section{Solution dynamics}
In this section, we discuss the dynamics of these general rogue wave
solutions.

To obtain the first-order rogue wave, we set $N=1$ in Theorem 1. In
this case,
\[
m_{11}^{(0)}=(x-2it-\frac{1}{2}+a_1)(x+2it-\frac{1}{2}+\bar
a_1)+\frac{1}{4},
\]
\[
m_{11}^{(1)}=(x-2it+\frac{1}{2}+a_1)(x+2it-\frac{3}{2}+\bar
a_1)+\frac{1}{4},
\]
hence the first-order rogue wave is
\begin{equation}  \label{f:uxtfirstorder}
u(x,t)=\frac{m_{11}^{(1)}}{m_{11}^{(0)}}
=\frac{(x-2it+\frac{1}{2}+a_1)(x+2it-\frac{3}{2}+\bar
a_1)+\frac{1}{4}} {(x-2it-\frac{1}{2}+a_1)(x+2it-\frac{1}{2}+\bar
a_1)+\frac{1}{4}}.
\end{equation}
Clearly, the complex parameter $a_1$ in this solution can be
normalized to zero by a shift of $x$ and $t$, as we have mentioned
before. After setting $a_1=0$, this first-order rogue wave can be
rewritten as
\begin{equation}  \label{e:1strogue}
u(x,t)=1-\frac{4(1-4it)}{1+4\hat{x}^2+16t^2},
\end{equation}
where $\hat{x}=x-1/2$. This rogue wave was first obtained by
Peregrine \cite{Rogue1}, see also \cite{Akhmediev_PRE}. Its maximum
peak amplitude is equal to 3, i.e., three times the background
amplitude.

To obtain the second-order rogue waves, we take $N=2$. In this case,
\begin{equation} \label{e:u2a}
u=\frac{\left|\matrix{ m_{11}^{(1)} &m_{13}^{(1)} \cr m_{31}^{(1)}
&m_{33}^{(1)}}\right|} {\left|\matrix{ m_{11}^{(0)} &m_{13}^{(0)}
\cr m_{31}^{(0)} &m_{33}^{(0)}}\right|}.
\end{equation}
{}From the previous discussions, we will set $a_1=a_2=0$. Then the
general second-order rogue wave can be obtained from (\ref{e:u2a})
as
\begin{equation}
u=1+\frac{\phi}{\psi},
\end{equation}
where
\begin{eqnarray}
\phi & = & 24\{(3 x-6 x^2+4 x^3-2 x^4-48
t^2+48 x t^2-48 x^2 t^2-160 t^4)\nonumber \\
&& +it(-12+12x-16 x^3+8 x^4 +32 t^2-64 xt^2 +64 x^2t^2 +128 t^4)
\nonumber \\
&& +6a_3(1-2 x+x^2-4 i t+4 i x t-4t^2) +6\bar{a}_3(- x^2+4 i x t+4
t^2)\},   \nonumber
\end{eqnarray}
\begin{eqnarray}
\psi& = & (9-36 x+72 x^2-72 x^3+72 x^4-48 x^5+16 x^6)
\nonumber \\
&& +96t^2(3+3x-4x^3+2x^4)+384t^4(5-2x+2x^2)+1024 t^6 \nonumber
\\ && +24(a_3+\bar{a}_3)(3x^2-2x^3-12t^2+24xt^2) +
48i(a_3-\bar{a}_3)(3t+6xt-6x^2t+8t^3)+144 a_3 \bar{a}_3,   \nonumber
\end{eqnarray}
and $a_3$ is a free complex parameter. We have found that the
maximum of $|u(x,t,a_3)|$ is equal to 5, and it is obtained when
\[a_3=-1/12. \]
At this $a_3$ value, the solution is
\begin{equation}
u_m(x,t)=1+\frac{\phi_m}{\psi_m},
\end{equation}
where
\[
\phi_m=9-72 \hat{x}^2-48 \hat{x}^4-864 t^2-3840 t^4-1152
\hat{x}^2t^2+it(-180-288 \hat{x}^2+192 \hat{x}^4+384 t^2+3072
t^4+1536 \hat{x}^2t^2),
\]
\[
\psi_m=\frac{9}{4}+27 \hat{x}^2+12\hat{x}^4+16 \hat{x}^6+396t^2+1728
t^4+1024 t^6-288\hat{x}^2t^2 +768 \hat{x}^2t^4 +192 \hat{x}^4t^2,
\]
and $\hat{x}=x-0.5$. This solution is displayed in Fig. 1(a). It is
easy to see that this solution is the special second-order rogue
wave obtained by Akhmediev et al. \cite{Akhmediev_PRE} (after a
shift in $x$). Thus the special second-order rogue wave obtained by
Akhmediev et al. is the one with the highest peak amplitude among
all second-order rogue waves. At other $a_3$ values, however, we can
obtain rogue waves which have very different solution dynamics from
that in Fig. 1(a). For instance, rogue waves at $a_3=5/3, -5i/2$ and
$5i/2$ are displayed in Fig. 1(b,c,d) respectively. In each of these
solutions, three intensity humps appear at different times and/or
space, and each intensity hump is roughly a first-order (Peregrine)
rogue wave (\ref{e:1strogue}). Specifically, in Fig. 1(b), the
solution features double temporal bumps (elevations) at $x\approx
-0.5$ and a single temporal bump at $x\approx 2.2$. In Fig. 1(c),
the solution first rises up and reaches a peak at $(x, t) \approx
(0.5, -0.7)$. Afterwards, the solution temporally decays at
$x\approx 0.5$, but two new bumps rise at the two sides. In Fig.
1(d), the solution is similar to that in Fig. 1(c) but with a time
reversal. Obviously, the rogue-wave dynamics in Fig. 1(b-d) are
quite different from the one in Fig. 1(a). The solution dynamics in
Fig. 1(b-d) resemble those reported in
\cite{Rogue_higher_order,Rogue_higher_order2,Rogue_Gaillard}.

\begin{figure}[h!]
\centerline{\includegraphics[width=0.5\textwidth]{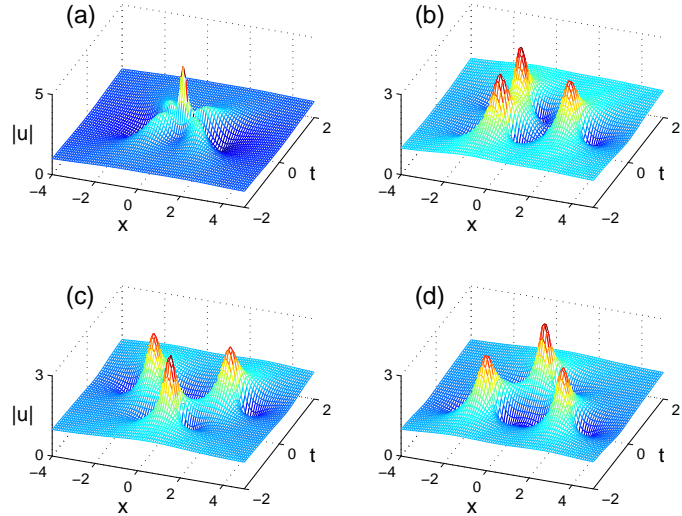}}
\caption{Second-order rogue waves with parameters: (a) $a_3=-1/12$;
(b) $a_3=5/3$; (c) $a_3=-5i/2$; (d) $a_3=5i/2$.  }
\end{figure}

Next we examine third-order rogue waves. In this case, we set
$a_1=a_2=a_4=0$ without loss of generality. If one takes
\[
a_3=-1/12,\quad a_5=-1/240,
\]
then the corresponding solution $u_m(x,t)$ is equal to the
third-order rogue wave obtained by Akhmediev et al.
\cite{Akhmediev_PRE} except a shift in $x$. This solution is
displayed in Fig. 2(a). The maximum amplitude of this solution is
equal to 7, which occurs at $(x,t)=(1/2,0)$. We have found that this
special rogue-wave solution $u_m(x,t)$ is also the one with the
highest peak amplitude among all third-order rogue waves $u(x,t;
a_3, a_5)$. But if we take other $(a_3, a_5)$ values, rogue waves
with dynamics different from Fig. 2(a) will be obtained. Three of
such solutions, with $(a_3, a_5)=(25/3, 0)$, $(-25i/3, 0)$ and $(0,
50i/3)$, are displayed in Fig. 2(b,c,d) respectively. These
solutions feature six intensity humps which appear at different
times and/or space, and each intensity hump is roughly a first-order
rogue wave (\ref{e:1strogue}). In Fig. 2(b), the solution exhibits
triple temporal bumps at $x\approx -2$, double temporal bumps at
$x\approx 2$, and a single temporal bump at $x\approx 6$. In Fig.
2(c), the solution develops a single hump first. Then this hump
decays, but two new humps rise simultaneously at the two sides. Then
these two humps decay, but three additional humps develop
simultaneously. In Fig. 2(d), two intensity humps rise
simultaneously at different spatial locations first. After they
decay, additional four intensity humps arise at different locations
and times. A remarkable feature in the rogue waves in Fig. 2(b-d) is
the high regularity of their spatiotemporal patterns. For instance,
the pattern in Fig. 2(c) is a highly symmetric triangle, while the
one in Fig. 2(d) is like a pentagon. These spatiotemporal patterns
of rogue waves are different from the ones reported in
\cite{Rogue_higher_order2,Rogue_Gaillard}.

\begin{figure}[h!]
\centerline{\includegraphics[width=0.5\textwidth]{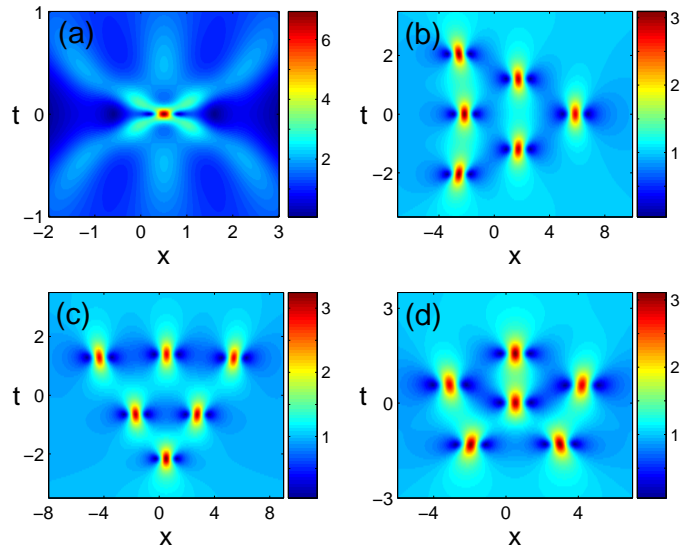}}
\caption{Third-order rogue waves with parameters $(a_3,a_5)$ as: (a)
$(-1/12, -1/240)$; (b) $(25/3, 0)$; (c) $(-25i/3, 0)$; (d) $(0,
50i/3)$. }
\end{figure}

The results shown above apparently can be extended to fourth- and
higher-order rogue waves. By special choices of the free parameters
$(a_3, a_5, a_7, \cdots)$, we can reproduce the rogue waves obtained
in \cite{Akhmediev_PRE} as special cases. But other choices of those
parameters can yield even richer spatiotemporal patterns, such as
triangular patterns like Fig. 2(c) but with more intensity humps
such as 10, 15, and so on.

\section{Summary and discussion}
In this paper, we derived general $N$-th order rogue waves in the
NLS equation by the bilinear method. These solutions were obtained
from Gram determinant solutions of bilinear equations through
dimension reduction and then further simplified to a very explicit
form. We showed that these general rogue waves contain $N-1$ free
irreducible complex parameters. By different choices of these free
parameters, we obtained rogue waves with novel spatiotemporal
patterns. These new spatiotemporal patterns reveal the rich dynamics
in rogue-wave solutions and deepen our understanding of the
rogue-wave phenomena. We also showed that the rogue waves reported
in \cite{Akhmediev_PRE} are special solutions with the highest peak
amplitude among all rogue waves of the same order.

We would like to point out that the new spatiotemporal patterns of
rogue waves obtained in this paper may also find applications in
other branches of applied mathematics and physics. For instance, the
triangular rogue-wave patterns in Figs. 1(c), 2(c) and their
higher-order extensions (with more intensity humps) closely resemble
the spike pattern which forms after the point of gradient
catastrophe in the semiclassical (zero-dispersion) limit of the NLS
equation (see Fig. 1 in \cite{Tovbis}). The connection between these
exact rogue-wave solutions and the semiclassical-NLS patterns is an
interesting question which lies outside the scope of the present
article.

\section*{Acknowledgment}
The work of Y.O. is supported in part by
JSPS Grant-in-Aid for Scientific Research (B-19340031, S-19104002)
and for challenging Exploratory Research (22656026),
and the work of J.Y.
is supported in part by the Air Force Office of Scientific Research
(Grant USAF 9550-09-1-0228) and the National Science Foundation
(Grant DMS-0908167).

\begin{center}
\textbf{\large Appendix}
\end{center}

In this appendix, we determine the number of free parameters in the
rogue-wave solutions obtained in this paper. Since the solutions in
Theorem 1 are derived from the ones in Lemma 2, we will examine
solutions in Lemma 2 below.

First we can factor out $a_0$ from $A_i$ and $b_0$ from $B_j$. These
factors cancel out in the formula (\ref{e:ugf}), thus we will set
$a_0=b_0=1$ without loss of any generality.

Secondly, let us consider the effect of a constant shifting of
$(x_1,x_2)$. By the shifting $(x_1,x_2)\to(x_1+\alpha,x_2+\beta)$,
$m^{(n)}$ in (\ref{e:mn}) gets an exponential factor,
\begin{equation}
m^{(n)}\to m^{(n)}e^{\theta}, \quad
\theta=(p+q)\alpha+(p^2-q^2)\beta,
\end{equation}
consequently the $(i,j)$-component $A_iB_jm^{(n)}$ in (\ref{e:mij})
is also modified. Below we show that $A_iB_jm^{(n)}$ is modified as
\begin{equation}  \label{e:ABappendix}
A_iB_jm^{(n)}\to A_iB_j(m^{(n)}e^{\theta})=e^{\theta}\hat A_i\hat
B_jm^{(n)},
\end{equation}
where
\begin{equation}
\hat A_i=\sum_{k=0}^i\frac{\hat a_k}{(i-k)!}(p\partial_p)^{i-k},
\quad \hat B_j=\sum_{l=0}^j\frac{\hat
b_l}{(j-l)!}(q\partial_q)^{j-l},
\end{equation}
and
\begin{eqnarray}
&&\hat a_k=\sum_{\nu=0}^ka_{\nu}S_{k-\nu}(\mbox{\boldmath
$x$}_0^{+}), \quad \mbox{\boldmath $x$}_0^{+}
=\left(p\alpha+2p^2\beta,\frac{p\alpha+4p^2\beta}{2},\cdots,
\frac{p\alpha+2^{k}p^2\beta}{k!},\cdots\right), \label{e:appendexak}
\\
&&\hat b_l=\sum_{\nu=0}^lb_{\nu}S_{l-\nu}(\mbox{\boldmath
$x$}_0^{-}), \quad \mbox{\boldmath $x$}_0^{-}
=\left(q\alpha-2q^2\beta,\frac{q\alpha-4q^2\beta}{2},\cdots,
\frac{q\alpha-2^{k}q^2\beta}{k!},\cdots\right). \label{e:appendexbk}
\end{eqnarray}
To prove (\ref{e:ABappendix}), we notice that for the generator
${\cal G}$ of the differential operators $(p\partial_p)^k$ defined
by
\begin{equation} \label{e:q3}
{\cal G}=\sum_{k=0}^{\infty}\frac{\lambda^k}{k!}(p\partial_p)^k
=\exp(\lambda p\partial_p)=\exp(\lambda \partial_{\ln p}),
\end{equation}
the relation
\[{\cal G}F(p,q)=F(e^{\lambda}p,q)\]
holds for any function $F(p,q)$. This relation is a special case of
the previous relation (\ref{r:GFpq}). Thus,
\[
e^{-\theta}{\cal G}(e^{\theta}F)
=\exp\left((e^{\lambda}-1)p\alpha+(e^{2\lambda}-1)p^2\beta\right){\cal
G}F
=\exp\left(\sum_{k=1}^{\infty}\frac{\lambda^k}{k!}(p\alpha+2^kp^2\beta)\right)
{\cal G}F,
\]
whose coefficient of order $\lambda^k$ gives
\[
\frac{1}{k!}(p\partial_p)^k(e^{\theta}F)
=e^{\theta}\sum_{\nu=0}^kS_{\nu}(\mbox{\boldmath $x$}_0^{+})
\frac{1}{(k-\nu)!}(p\partial_p)^{k-\nu}F.
\]
Similarly we have
\[
\frac{1}{l!}(q\partial_q)^l(e^{\theta}F)
=e^{\theta}\sum_{\nu=0}^lS_{\nu}(\mbox{\boldmath $x$}_0^{-})
\frac{1}{(l-\nu)!}(q\partial_q)^{l-\nu}F.
\]
Therefore,
\begin{eqnarray*}
&&A_iB_j(m^{(n)}e^{\theta}) =\sum_{k=0}^i\sum_{l=0}^ja_kb_l
\frac{1}{(i-k)!}(p\partial_p)^{i-k}
\frac{1}{(j-l)!}(q\partial_q)^{j-l}(m^{(n)}e^{\theta})
\\
&&\quad =e^{\theta}\sum_{k=0}^i\sum_{l=0}^ja_kb_l
\sum_{\mu=0}^{i-k}S_{\mu}(\mbox{\boldmath $x$}_0^{+})
\frac{1}{(i-k-\mu)!}(p\partial_p)^{i-k-\mu}
\sum_{\nu=0}^{j-l}S_{\nu}(\mbox{\boldmath $x$}_0^{-})
\frac{1}{(j-l-\nu)!}(q\partial_q)^{j-l-\nu}m^{(n)}
\\
&&\quad =e^{\theta}\sum_{k=0}^i\sum_{l=0}^j \frac{\hat
a_k}{(i-k)!}(p\partial_p)^{i-k} \frac{\hat
b_l}{(j-l)!}(q\partial_q)^{j-l}m^{(n)} =e^{\theta}\hat A_i\hat
B_jm^{(n)},
\end{eqnarray*}
which proves Eq. (\ref{e:ABappendix}).

Now we take $p=q=1$. Then from Eqs.
(\ref{e:appendexak})-(\ref{e:appendexbk}), we get
\[
\hat a_0=a_0=1, \quad \hat a_1=a_1+\alpha+2\beta, \quad \cdots,
\quad \hat b_0=b_0=1, \quad \hat b_1=b_1+\alpha-2\beta, \quad
\cdots.
\]
Thus by a shifting of $(x_1,x_2)\to(x_1+\alpha,x_2+\beta)$ with
$\alpha=-(a_1+b_1)/2$ and $\beta=-(a_1-b_1)/4$, we obtain $\hat
a_1=\hat b_1=0$. When this shifting is combined with shifts of
higher coefficients $a_2\to\hat a_2$, $a_3\to\hat a_3$, $\cdots$,
$b_2\to\hat b_2$, $b_3\to\hat b_3$, $\cdots$, the solution $\tau_n$
depends on parameters $(\hat a_2,\hat a_3,\cdots;\hat b_2,\hat
b_3,\cdots)$ only. In other words, by a shift of $(x_1,x_2)$, we can
normalize $a_1=b_1=0$.

Thirdly, from the expressions of $m^n_{ij}$ in (\ref{e:mij}) and the
expressions of $A_i$ and $B_j$, we see that in the determinant
formula for $\tau_n$ in (\ref{e:tau}), when we subtract the product
of the first row and $a_2$ from the second row, and subtract the
product of the second row and $a_2$ from the third row, \dots, and
subtract the product of the $i$th row and $a_2$ from the $(i+1)$-th
row, and then subtract the product of the first column and $b_2$
from the second column, and subtract the product of the second
column and $b_2$ from the third column, etc., we can remove the
parameter $a_2$ and $b_2$ from the solution formula (\ref{e:tau}).
By similar treatments, we can remove all other even coefficients
$a_4, a_6, \cdots$ and $b_4, b_6, \cdots$ as well. In other words,
we can set $a_2=a_4=a_6=\cdots=0$ and $b_2=b_4=b_6=\cdots=0$ without
any loss of generality.

By summarizing the above results, we see that without any loss of
generality, we can set
\[
a_0=b_0=1, \quad a_2=a_4=a_6=\cdots=b_2=b_4=b_6=\cdots=0.
\]
In addition, by a shift of $(x_1,x_2)$, we can normalize
$a_1=b_1=0$. Combined with the complex conjugacy condition $b_k=\bar
a_k$ in (\ref{e:ccab}), we then find that the $N$-th order
rogue-wave solutions in Theorem 1 have $N-1$ free irreducible
complex parameters, $a_3, a_5, \dots, a_{2N-1}$.


\begin{thebibliography}O

\bibitem{Rogue_nature1} D. R. Solli, C. Ropers, P. Koonath and B. Jalali,
``Optical rogue waves", Nature 450, 1054-1057 (2007).

\bibitem{Rogue_nature2}
B. Kibler, J. Fatome, C. Finot, G. Millot, F. Dias, G. Genty, N.
Akhmediev, J.M. Dudley, ``The Peregrine soliton in nonlinear fibre
optics", Nature Physics, 6, 790-795 (2010).


\bibitem{Benney}
D.J. Benney and A.C. Newell, ``Nonlinear wave envelopes", J. Math.
Phys. 46, 133 (1967).

\bibitem{Zakharov}
V.E. Zakharov, ``Stability of periodic waves of finite amplitude on
the surface of a deep fluid," J. Appl. Mech. Tech. Phys. 9, 190-94
(1968).

\bibitem{Hasegawa}
A. Hasegawa and F. Tappert, ``Transmission of stationary nonlinear
optical pulses in dispersive dielectric fibers", Appl. Phys. Lett.
23, 142 (1973).

\bibitem{ZS}
V.E. Zakharov and A.B. Shabat, ``Exact theory of two-dimensional
self-focusing and one-dimensional self-modulation of waves in
nonlinear media", Sov. Phys. JETP 34, 62 (1972).

\bibitem{Rogue1}
D. H. Peregrine, ``Water waves, nonlinear Schr\"odinger equations
and their solutions," J. Australian Math. Soc. B, 25, 16–-43 (1983).

\bibitem{Akhmediev_PRE}
N. Akhmediev, A. Ankiewicz, and J. M. Soto-Crespo, ``Rogue Waves and
Rational Solutions of the Nonlinear Schrödinger Equation," Phys.
Rev. E 80, 026601 (2009).

\bibitem{Rogue_higher_order}
P. Dubard, P. Gaillard, C. Klein, V.B. Matveev, ``On multi-rogue
wave solutions of the NLS equation and positon solutions of the KdV
equation", Eur. Phys. J. Special Topics 185, 247–258 (2010).

\bibitem{Rogue_higher_order2}
P. Dubard, V.B. Matveev, ``Multi-rogue waves solutions to the
focusing NLS equation and the KP-I equation", Nat. Hazards. Earth.
Syst. Sci. 11, 667-672 (2011).

\bibitem{Rogue_Gaillard}
P. Gaillard, ``Families of quasi-rational solutions of the NLS
equation and multi-rogue waves", J. Phys. A: Math. Theor. 44, 435204
(2011).


\bibitem{Ablowitz_homo}
M.J. Ablowitz and B.M Herbst, ``On homoclinic structure and
numerically induced chaos for the nonlinear Schr6dinger equation",
SIAM J. Appl. Math. 50, 339 (1990).

\bibitem{Rogue_homo}
N. Akhmediev, A. Ankiewicz and M. Taki, ``Waves that appear from
nowhere and disappear without a trace", Phys. Lett. A 373, 675-678
(2009).

\bibitem{Akhmediev_breather}
N. Akhmediev, J. M. Soto-Crespo and A. Ankiewicz, ``Extreme waves
that appear from nowhere: on the nature of rogue waves", Phys. Lett.
A, 373 2137-2145 (2009).

\bibitem{H}
R. Hirota, \emph{The direct method in soliton theory} (Cambridge
University Press, Cambridge, 2004).

\bibitem{S}
M. Sato,
``Soliton equations as dynamical systems on a infinite dimensional
Grassmann manifolds",
RIMS Kokyuroku, 439, 30 (1981).

\bibitem{JM}
M. Jimbo and T. Miwa,
``Solitons and infinite dimensional Lie algebras",
Publ. RIMS, Kyoto Univ., 19, 943 (1983).

\bibitem{IY}
T. Ikeda and H.-F. Yamada,
``Polynomial $\tau$-functions of the NLS-Toda hierarchy and
the Virasoro singular vectors",
Lett. Math. Phys., 60, 147 (2002).

\bibitem{Tovbis}
M. Bertola and A. Tovbis, ``Universality for the focusing nonlinear
Schr\"odinger equation at the gradient catastrophe point: Rational
breathers and poles of the tritronquee solution to Painleve I",
arXiv:1004.1828 (2010).

\end{thebibliography}
\end{document}